\documentclass{emulateapj}
\usepackage{apjfonts}

\begin{document}
\slugcomment{Publications of the Astronomical Society of the Pacific, ACCEPTED}

\shorttitle{THE PRODUCTION AND EMPLOYMENT OF ASTRONOMERS}
\shortauthors{METCALFE}
 
\title{The Production Rate and Employment of Ph.D.\ Astronomers}

\author{Travis~S. Metcalfe}

\affil{High Altitude Observatory and Scientific Computing Division,\\ 
National Center for Atmospheric Research, P.O.\ Box 3000, Boulder, CO 
80307 USA}
\email{travis@hao.ucar.edu}

\begin{abstract}
In an effort to encourage self-regulation of the astronomy job market, 
I examine the supply of, and demand for, astronomers over time. On the 
supply side, I document the production rate of Ph.D.\ astronomers from 
1970 to 2006 using the UMI Dissertation Abstracts database, along with 
data from other independent sources. I compare the long-term trends in 
Ph.D.\ production with federal astronomy research funding over the same 
time period, and I demonstrate that additional funding is correlated with 
higher subsequent Ph.D.\ production. On the demand side, I monitor the 
changing patterns of employment using statistics about the number and 
types of jobs advertised in the AAS Job Register from 1984 to 2006. 
Finally, I assess the sustainability of the job market by normalizing this 
demand by the annual Ph.D.\ production. The most recent data suggest that 
there are now annual advertisements for about one postdoctoral job, half a 
faculty job, and half a research/support position for every new domestic 
Ph.D.\ recipient in astronomy and astrophysics. The average new astronomer 
might expect to hold up to 3 jobs before finding a steady position.
\end{abstract}

\keywords{sociology of astronomy}

\section{Background}

From 1982 to 1992, federal funding for astronomy research increased by 
more than 80\%, fueled primarily by NASA funding related to the Hubble 
Space Telescope. Trailing this trend by several years, the annual 
production of new Ph.D.\ astronomers more than doubled. Meanwhile, the 
total number of jobs advertised in the American Astronomical Society (AAS) 
Job Register remained fairly constant. Responsible undergraduate programs 
began to lecture incoming astronomy majors about this imbalance in the job 
market, comparing the statistical odds of a long-term career in astronomy 
to the chances of becoming a professional athlete. The faculty in some 
graduate programs started to debate the idea of limiting the number of 
incoming students (``birth control''), and they made efforts to track and 
publicize the long-term career progress of their Ph.D.\ recipients (e.g., 
see Dinerstein 1996).

It was in this atmosphere that Thronson (1991) devised a model to describe 
the surplus production of new astronomers. He observed that overproduction 
appears to be built into the system, making the mathematical formulation 
of the problem similar to that of industrial pollution---an unintended 
side effect of the process. He calculated the future astronomer surplus 
under a variety of possible circumstances, including expected retirement 
rates and increased research funding. The most likely of the scenarios 
predicted a short-lived reduction in the astronomer surplus over a decade 
or so, followed by continued growth in the overproduction rate.

A few years later, Harris (1994) identified some reliable sources of data 
to evaluate the astronomy job market over time. He looked at the number of 
dissertations published in the subject area of astronomy and astrophysics 
using the UMI Dissertation Abstracts database (UMI) to track new Ph.D.\ 
recipients. He noted that the American Institute of Physics (AIP) 
tabulates the number of degrees awarded by astronomy departments annually, 
but astronomers who receive their degrees from physics departments have 
not been tracked until the past few years. To quantify the number of 
long-term jobs available each year, Harris examined advertisements in the 
AAS Job Register. This has some inherent limitations, but it is still the 
best source available.

Thanks to the Internet, most of these data sources are now much more 
accessible, and we can easily examine the trends in Ph.D.\ production and 
job availability over an extended time baseline. Further data on the 
trends in astronomy research funding are also available, allowing us to 
investigate the possible connection with Ph.D.\ production predicted by 
Thronson (1991). In section~\ref{DATA}, I document the sources of these 
data for future reference and I present tables that can be updated 
annually as new statistics become available. I examine the historical 
trends of both the supply and demand in section~\ref{RESULTS}, and I 
discuss the results in section~\ref{DISC}.

\section{Data Sources\label{DATA}}

The Statistical Research Center of the AIP has been surveying every 
astronomy Ph.D.\ program in the U.S.\ since 1972. An archive of their 
annual reports beginning in 1980 is available online\footnote{ 
http://aip.org/statistics/trends/archives/astrorost.htm}, and the most 
recent ``Roster of Astronomy Departments'' (Nicholson \& Mulvey 2007) 
appears on a separate 
page\footnote{http://aip.org/statistics/trends/gradtrends.html}. I 
received the complete set of data on astronomy Ph.D.\ production from AIP 
staff (P.~Mulvey, personal communication), which is listed under the 
heading ``AIP'' in Table~\ref{tab1}. These totals do not include degrees 
that were conferred by physics departments to Ph.D.\ recipients who 
complete a dissertation in a subject related to astrophysics, even though 
this is the most popular subfield among domestic first year physics 
graduate students (Mulvey \& Tesfaye 2006).

\begin{table}[t]
\begin{center}
\caption{Ph.D.\ Production and Research Funding.\label{tab1}}
\begin{tabular}{lrrrrrr}
\hline\hline
 & \multicolumn{3}{c}{Ph.D.s} & 
\multicolumn{3}{c}{Research Funding (M\$)} \\
Year & AIP    & SED    & UMI & Total   & NASA    & NSF     \\
\hline
1970 &$\cdots$& 111    & 141 & 136.455 & 100.875 &  19.922 \\
1971 &$\cdots$& 113    & 119 & 126.852 &  90.688 &  23.236 \\
1972 &  75    & 129    & 173 & 136.405 &  89.859 &  25.608 \\
1973 &  80    & 131    & 147 & 123.085 &  82.020 &  27.080 \\
1974 &  97    & 133    & 172 & 136.030 &  91.564 &  30.283 \\
1975 &  76    & 131    & 162 & 168.319 & 122.002 &  29.093 \\
1976 &  93    & 150    & 186 & 163.093 & 116.575 &  31.754 \\
1977 &  96    & 120    & 164 & 197.305 & 140.702 &  41.585 \\
1978 &  93    & 138    & 173 & 213.601 & 148.603 &  46.614 \\
1979 &  90    & 115    & 152 & 284.429 & 213.272 &  51.064 \\
1980 & 106    & 121    & 133 & 285.671 & 229.724 &  34.512 \\
1981 &  82    & 109    & 144 & 281.027 & 198.674 &  59.146 \\
1982 &  70    & 102    & 121 & 275.784 & 184.630 &  65.411 \\
1983 &  81    & 115    & 129 & 357.416 & 265.687 &  64.638 \\
1984 &  74    &  98    & 120 & 382.292 & 277.013 &  77.534 \\
1985 &  66    & 100    & 150 & 414.743 & 310.668 &  76.554 \\
1986 &  86    & 109    & 132 & 468.160 & 362.547 &  77.808 \\
1987 &  72    & 100    & 166 & 522.187 & 410.651 &  76.789 \\
1988 &  94    & 130    & 171 & 470.838 & 364.730 &  77.638 \\
1989 &  94    & 113    & 223 & 541.939 & 428.963 &  83.373 \\
1990 &  89    & 128    & 240 & 597.154 & 462.320 & 104.613 \\
1991 &  73    & 125    & 233 & 631.657 & 501.766 & 101.712 \\
1992 &  93    & 134    & 254 & 739.057 & 584.504 & 116.126 \\
1993 & 119    & 145    & 256 & 686.281 & 520.184 & 109.711 \\
1994 & 117    & 144    & 271 & 747.590 & 562.962 & 122.065 \\
1995 & 133    & 173    & 321 & 764.458 & 593.089 & 126.225 \\
1996 & 126    & 192    & 303 & 728.958 & 570.450 & 119.946 \\
1997 & 117    & 198    & 320 & 774.740 & 621.324 & 115.212 \\
1998 & 116    & 206    & 310 & 732.150 & 575.093 & 113.248 \\
1999 &  88    & 159    & 267 & 757.859 & 578.973 & 124.395 \\
2000 & 139    & 185    & 291 & 880.267 & 710.006 & 130.702 \\
2001 & 101    & 186    & 253 & 759.195 & 559.539 & 153.919 \\
2002 & 102    & 141    & 251 & 751.527 & 540.243 & 167.921 \\
2003 &  88    & 167    & 254 & 871.757 & 639.761 & 192.179 \\
2004 & 116    & 165    & 297 & 920.763 & 678.304 & 202.092 \\
2005 &  91    & 186    & 269 &$\cdots$ &$\cdots$ &$\cdots$ \\
2006 & 119    &$\cdots$& 284 &$\cdots$ &$\cdots$ &$\cdots$ \\
\hline\hline
\end{tabular}
\tablecomments{Degree data for astronomy departments from the
American Institute of Physics (AIP), for astronomy and physics
departments from the Survey of Earned Doctorates (SED), and for
all astrophysics related dissertations from the UMI Dissertation
Abstracts database (UMI). Funding data for all R\&D (Total) with
the contributions from NASA and the NSF, reported as actual
dollar amounts (not adjusted for inflation).}
\end{center}
\end{table}

The NSF Division of Science Resources Statistics has documented the annual 
number of astronomy and astrophysics Ph.D.\ recipients through the Survey 
of Earned Doctorates (SED), a voluntary program administered by graduate 
deans at accredited universities. Over the past decade, more than 90\% of 
new doctorates have participated in this survey---at many institutions, it 
is now a requirement for graduation. Historical data from this survey were 
compiled in a recent NSF report (Thurgood et al.~2006), while data from 
the past decade appeared in a separate publication (Hill 2007). The data 
from 1970 to 2005, which include degrees conferred by both astronomy and 
physics departments, are listed under the heading ``SED'' in 
Table~\ref{tab1}.

\begin{table}[t]
\begin{center}
\caption{Number of AAS Job Register Ads.\label{tab2}}
\begin{tabular}{lrrrrrrrc}
\hline\hline
Year & All & PV     & TT     & NT     & R      & RS     & MO     & F      \\
\hline
1984 & 275 &  77    &  66    &$\cdots$&  72    &$\cdots$&$\cdots$&$\cdots$\\ 
1985 & 338 & 149    &  81    &$\cdots$& 118    &$\cdots$&$\cdots$&$\cdots$\\ 
1986 & 257 &$\cdots$&$\cdots$&$\cdots$&$\cdots$&$\cdots$&$\cdots$&$\cdots$\\ 
1987 & 240 &$\cdots$&$\cdots$&$\cdots$&$\cdots$&$\cdots$&$\cdots$&$\cdots$\\ 
1988 & 321 & 157    &$\cdots$&$\cdots$&$\cdots$&$\cdots$&$\cdots$&$\cdots$\\ 
1989 & 351 &$\cdots$&$\cdots$&$\cdots$&$\cdots$&$\cdots$&$\cdots$&$\cdots$\\ 
1990 & 240 & 120    &  32    &$\cdots$&$\cdots$&$\cdots$&$\cdots$&$\cdots$\\ 
1991 & 304 & 137    &  30    &$\cdots$&$\cdots$&$\cdots$&$\cdots$&$\cdots$\\ 
1992 & 327 & 127    &  40    & 13     &  63    & 57     &  21    & 0.140  \\
1993 & 315 &  66    &  59    & 12     &  42    & 60     &  23    & 0.160  \\ 
1994 & 318 & 140    &  44    & 15     &  14    & 76     &  24    & 0.233  \\ 
1995 & 425 & 200    &  54    & 23     &  73    & 52     &  17    & 0.268  \\ 
1996 & 426 & 210    &  63    & 15     &  65    & 56     &  15    & 0.211  \\ 
1997 & 469 & 208    &  74    &  9     &  55    & 49     &  65    & 0.230  \\ 
1998 & 569 & 224    &  77    & 23     & 112    & 68     &  51    & 0.210  \\ 
1999 & 623 & 206    &  94    & 43     & 130    & 38     &  67    & 0.230  \\ 
2000 & 739 & 277    & 108    & 39     & 132    & 66     &  63    & 0.264  \\ 
2001 & 715 & 310    &  93    & 59     &  93    & 56     &  78    & 0.288  \\ 
2002 & 716 & 370    & 117    & 29     & 109    &  3     &  46    & 0.198  \\ 
2003 & 722 & 389    & 119    & 22     &  78    & 29     &  85    & 0.072  \\ 
2004 & 714 & 333    & 113    & 22     & 117    & 28     & 101    & 0.251  \\ 
2005 & 673 & 334    & 111    & 14     &  88    & 19     & 102    & 0.276  \\ 
2006 & 738 & 388    & 129    & 22     &  87    & 19     &  93    & 0.276  \\
\hline\hline
\end{tabular}
\tablecomments{The total number of AAS Job Register ads (All) along with
the number of Postdoctoral and Visitor (PV), Tenure-Track (TT), Non-Tenure
track (NT), Research (R), Research Support (RS), Management and Other (MO)
positions advertised, and the fraction of ads for Foreign (F) jobs.}
\end{center}
\end{table}

Although the SED data represent a substantially complete survey, they do 
not include dissertations that are relevant to astronomy and astrophysics 
but were completed in another primary field. Following Harris (1994), I 
used the online version of the UMI Dissertation Abstracts 
database\footnote{http://proquest.umi.com/login} to obtain an independent 
estimate of the number of dissertations related to astronomy and 
astrophysics that were published annually from 1970 to 2006. For each 
calendar year, I queried the ``Dissertations \& Theses'' database for 
doctoral dissertations with a subject code of astronomy or astrophysics 
that were written in English and awarded a Ph.D.\ by a domestic 
university. The results are shown in Table~\ref{tab1} under the heading 
``UMI''. Since these queries included dissertations with primary subject 
codes in other fields (e.g., mathematics or chemistry) the net effect is 
to slightly inflate the supply of Ph.D.\ recipients, leading to a worst 
case scenario for job seekers. While the normalization of job market 
numbers to assess sustainability may be sensitive to the choice of degree 
data, the overall {\it trends} should be similar.

In addition to degree data, the NSF Division of Science Resources 
Statistics also documents the trends in federal funding for research and 
development. One of their recent reports provides a breakdown of federal 
funding from 1970 to 2003 by both agency and the detailed field of science 
and engineering (Meeks 2004). I obtained data for the most recent years 
from several additional NSF reports (Meeks 2005, 2006; Jankowski 2007). 
Table~\ref{tab1} includes the total research funding, along with the 
amounts contributed by NASA and the NSF, reported as the actual dollar 
values (in millions) for each year\footnote{The specific programs included 
in these totals were determined by each agency in their response to the 
annual Survey of Federal Funds for Research and Development}. To correct 
these values for inflation, I used the composite outlay deflators for 
``other grants'' from the FY\,2006 U.S.\ 
budget\footnote{http://gpoaccess.gov/usbudget/fy06/sheets/hist10z1.xls}, 
which normalizes all values to constant FY\,2000 dollars. Unfortunately, 
even the historical values of the inflation indices adopted by the 
government appear to change slightly from one budget to the next. So, any 
future updates to the research funding data in Table~\ref{tab1} should 
re-adjust the historical numbers using deflators from the most recent 
federal budget.

\begin{figure}[t]
\includegraphics[height=\linewidth, angle=-90]{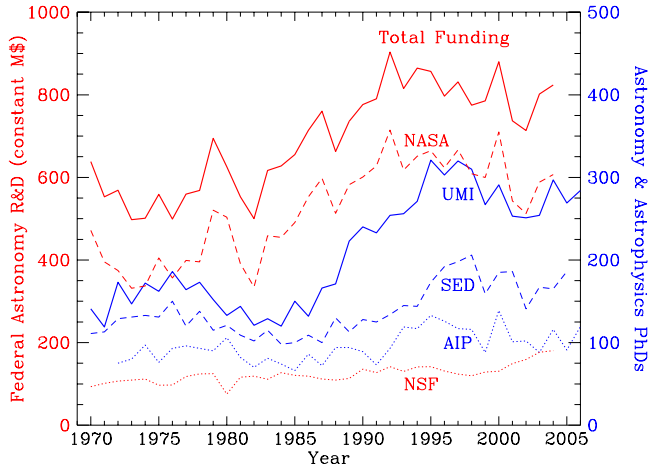}
\caption{Left axis: Total federal funding for astronomy research and 
development in millions of FY\,2000 dollars (upper solid line), including 
the annual totals from NASA (upper dashed line) and the NSF (lower dotted 
line). Right axis: Historical Ph.D.\ production from AIP surveys of 
astronomy departments (upper dotted line), from the NSF Survey of Earned 
Doctorates (SED; lower dashed line) and from all dissertations included in 
the UMI Dissertation Abstracts database and categorized in the subject 
area of astronomy and astrophysics (lower solid line).\label{fig1}}
\end{figure}

The total number of jobs advertised in the Job Register from 1984 to 2006 
is available from the AAS Annual Reports, published each year in the BAAS 
(e.g., Marvel 2007). A detailed breakdown by advertiser-specified job type 
has been compiled consistently since 1992, including the following 
categories: postdoctoral and visitor (PV), tenure-track (TT), 
non-tenure-track (NT), research (R), research support (RS), management and 
other (MO). The available data for each of these categories are listed in 
Table~\ref{tab2}. Note that the total number of jobs advertised in each 
year (shown under the heading ``All'') is not exactly equal to the sum of 
the six categories listed. The total also includes jobs categorized as 
predoctoral, which are omitted from Table~\ref{tab2} because they are not 
relevant to this study. Statistics about the type of institution offering 
the jobs are also compiled, including whether it is in a foreign country. 
To evaluate this contribution to the demand for astronomers, the final 
column of Table~\ref{tab2} lists the fraction of jobs that were classified 
as foreign (F). The primary limitation of the Job Register data is that 
they provide no way of evaluating which advertised positions are {\it 
new}, i.e. which jobs arise from retirement or death as opposed to faculty 
shuffle. This limitation is generally not relevant to shorter-term 
postdoctoral jobs, and assuming a reasonably constant number of faculty 
shuffles over time it will affect the normalization but not the trends in 
longer-term positions. This may partially offset the impact of adopting 
the slightly inflated degree data from UMI.


\section{Results\label{RESULTS}}

Using these data, drawn from reliable and accessible sources that are
likely to be updated annually, we can evaluate the current state of the 
astronomy job market and how it has changed over time. 

\begin{figure}[t]
\includegraphics[height=\linewidth, angle=-90]{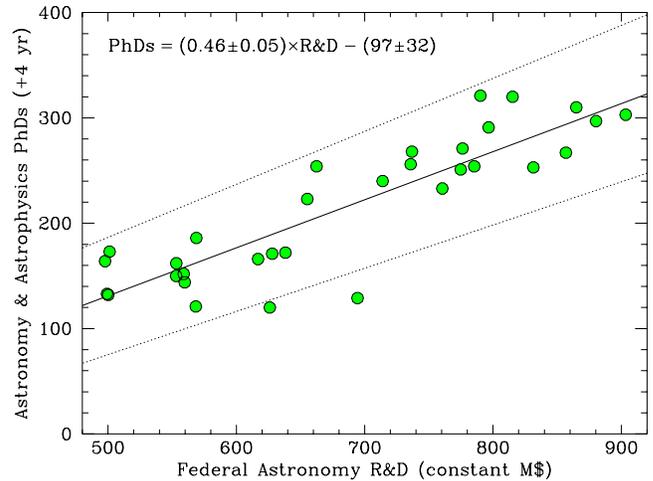}
\caption{The correlation between inflation-adjusted federal research 
funding and Ph.D.\ production related to astronomy and astrophysics 4 
years later. The best linear fit (solid line) suggests that an increase of 
\$100 million in federal funding ultimately leads to the production of 
40-50 additional Ph.D.\ recipients.\label{fig2}}
\end{figure}

\subsection{Supply of Astronomers}

The data from Table~\ref{tab1} on the supply of astronomers, and the 
inflation-corrected federal research funding that might be one of its 
primary drivers, are shown in Figure~\ref{fig1}. The left axis shows 
funding levels in millions of FY\,2000 dollars for the total federal 
investment (upper solid line), and for the contributions from NASA (upper 
dashed line) and the NSF (lower dotted line). The most striking thing 
about these data is how NASA dominates the total federal funding for 
astronomy research, and how it is responsible for nearly all of the 
volatility. By contrast, astronomy funding from the NSF has shown fairly 
steady growth, nearly doubling its real value over the last 35 years. 
Between 1982 and 1992, NASA funding for astronomy more than doubled. This 
expansion was driven primarily by preparations for the Hubble Space 
Telescope, but also by the Compton Gamma Ray Observatory and other 
missions (E.~Suckow, personal communication).

The right axis of Figure~\ref{fig1} shows the total number of Ph.D.\ 
recipients each year in astronomy and astrophysics from UMI (lower solid 
line) and SED (lower dashed line), as well as the subset of those coming 
from pure astronomy departments (upper dotted line). The correlation 
between the trend in Ph.D.\ recipients and the trend in federal funding is 
apparent, though there may be a delay of several years between the two 
curves. Degrees from astronomy departments account for about 25\% of the 
observed increase in Ph.D.\ recipients between the mid-1980's and 
mid-1990's, while most of the new degrees came from other departments.

To examine the relationship between federal funding and Ph.D.\ production 
more quantitatively, Figure~\ref{fig2} shows the correlation between 
inflation-adjusted research funding in a given year and the number of new 
degrees related to astronomy and astrophysics 4 years later (from UMI). 
Similar correlations were produced for time delays ranging from 0 to 5 
years, but a 4 year delay minimized the fractional errors on the fit 
parameters (formally, all of the fits were statistically equivalent). The 
best linear fit is shown as a solid line, with the $\pm1\sigma$ limits on 
the fit parameters indicated with dotted lines. The correlation suggests 
that an incremental increase of \$100 million in federal funding may lead 
to the production of 40-50 additional Ph.D.\ recipients in subsequent 
years (4 years is the median time to Ph.D.\ after the initial 2 years of 
coursework).

\begin{figure}[t]
\includegraphics[height=\linewidth, angle=-90]{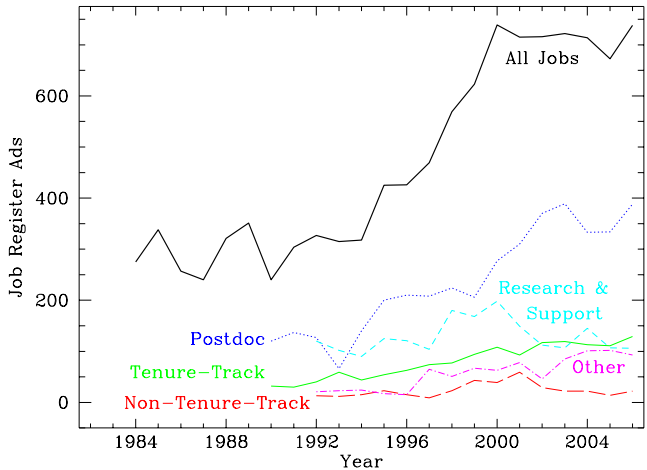}
\caption{The number of jobs advertised in the AAS Job Register from 1984 
to 2006 (upper solid line). A detailed breakdown by advertiser-specified 
job type is available beginning in the early 1990's, and is shown for 
positions categorized as postdoctoral and visitor (dotted line), research 
and research support (dashed line), tenure-track (lower solid line), 
non-tenure-track (long dashed line), along with management and other 
positions (dot-dashed line).\label{fig3}}
\end{figure}

\subsection{Demand for Astronomers}

The data from Table~\ref{tab2} on the demand for astronomers from 1984 to 
2006 are shown in Figure~\ref{fig3}. The total number of jobs advertised 
(upper solid line) is shown along with the numbers of postdoctoral and 
visitor positions (dotted line), research and research support jobs 
(dashed line), as well as tenure-track (lower solid line), 
non-tenure-track (long dashed line), management and other (dash-dot line) 
advertisements. The growth in the total number of jobs advertised 
throughout the 1990's is remarkable, and is dominated by an increase in 
the number of postdoctoral positions, though all job categories 
experienced some growth during this period. The most recent data show some 
anti-correlated trends after the year 2000, suggesting a possible shift 
from research and support positions to postdocs, and from non-tenure-track 
to tenure-track jobs.

\begin{figure}[t]
\includegraphics[height=\linewidth, angle=-90]{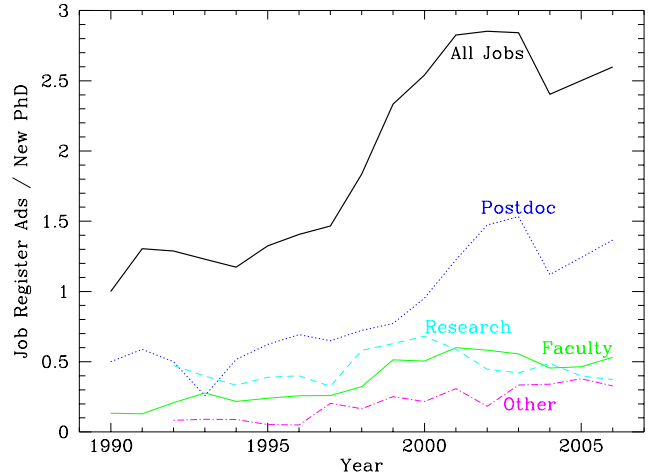}
\caption{Job advertisements in several categories, normalized by the UMI 
measure of astronomy and astrophysics Ph.D.\ production from 1990 to 2006. 
There is now more than one postdoctoral or visitor position (dotted line), 
about half a tenure-track or non-tenure-track job (lower solid line) and 
half a research or research support position (dashed line) advertised for 
each new Ph.D.\ recipient. If all advertised jobs (upper solid line) are 
filled each year, then the average new Ph.D.\ might expect to hold between 
2 and 3 of them before finding a stable position.\label{fig4}}
\end{figure}

To assess the relative sustainability of the job market, we can normalize 
these job advertisements by the number of new Ph.D.\ recipients in each 
year (from UMI). These data are shown in Figure~\ref{fig4} for all 
advertised jobs (upper solid line), postdoctoral and visitor positions 
(dotted line), research and research support jobs (dashed line), 
tenure-track and non-tenure-track positions (lower solid line), as well as 
management and other advertisements (dash-dot line). By normalizing in 
this way, the vertical axis can be thought of as a relative probability of 
obtaining a given type of position for each class of Ph.D.\ recipients. It 
is not exactly a probability, since the number of foreign Ph.D.\ 
recipients competing for some jobs may exceed the number of foreign jobs 
advertised, and because some Ph.D.\ recipients from previous years will be 
competing for some jobs while others will eventually choose to work 
outside the field or in a foreign country.

There now appears to be more than one postdoctoral position advertised for 
every new domestic Ph.D.\ recipient related to astronomy and astrophysics. 
This surge in the relative probability of obtaining a postdoc position can 
be traced to both a slightly declining population of new Ph.D.\ recipients 
since a peak in the mid-1990's (Figure~\ref{fig1}), as well as the larger 
number of postdoc positions advertised---possibly at the expense of 
longer-term research jobs (Figure~\ref{fig3}). At best, half of all recent 
Ph.D.\ recipients might expect to obtain a faculty position, while the 
other half will likely end up in a research or support position. While 
this situation may not be ideal for job seekers who prefer to remain in 
academia, it was much more competitive a decade ago with nearly 6 new 
Ph.D.\ astronomers in 1995 for every new tenure-track job. If we assume 
that all advertised jobs are eventually filled by someone in a related 
field, then the upper solid line in Figure~\ref{fig4} suggests that the 
average new Ph.D.\ might expect to hold between 2 and 3 jobs before 
finding a more permanent position.


\section{Discussion\label{DISC}}

We are now in a position to evaluate the predictions of Thronson (1991), 
who made 20 year projections of several models for the production of 
astronomers over time. His definition of the ``astronomer surplus'' was 
the annual ratio of new astronomy Ph.D.\ recipients tabulated by the AIP 
relative to new tenure-track faculty positions advertised in the AAS Job 
Register, which he calculated to be around 2.5 in 1991. Regardless of the 
absolute level of overproduction, his most realistic models included the 
competing effects of increased research funding (which leads to growth in 
the surplus), and retirement (which briefly absorbs some of the surplus). 
These models predicted a gradual decline in the astronomer surplus over 
about 10 years, followed by continued overproduction:{\it ``A momentary 
abundance of jobs will be subsequently compensated for by training of new 
graduate students by a new cohort of young professors''}. Indeed, the 
anticipated impact of increased research funding on Ph.D.\ production has 
now been quantified in Figure~\ref{fig2}. Furthermore, the slowly evolving 
ratio of faculty positions to new Ph.D.\ recipients in Figure~\ref{fig4} 
shows exactly the predicted behavior---peaking in 2001, and gradually 
leveling off or declining in more recent years.

Despite the accuracy of Thronson's predictions, a broader definition of 
sustainability may now be necessary, since the character of the astronomy 
job market appears to have shifted in the meantime. Astronomy research is 
increasingly moving towards automation and large collaborations where 
service positions are becoming more important (Sage 2001). This cultural 
shift might be reflected in the surge of new research and support 
positions between 1997 and 2002, which may have absorbed some of the 
astronomer surplus generated during the more competitive conditions of the 
early 1990's. More recently, as the new cadre of young faculty have 
attempted to recruit a limited supply of graduate students, some may have 
turned to postdocs to maintain their research productivity. This could be 
responsible for the unprecedented expansion of postdoctoral jobs after 
1999---but as startup funds are depleted and external research funding 
stays relatively constant, these ``holding pattern'' positions could 
evaporate. The most recent data show the first signs of a possible 
slowdown.

The most important aspect of the ongoing cultural shift in the astronomy 
job market is the persistent gap between expectations and reality. When 
first year graduate students are surveyed by the AIP, fully 87\% of those 
in astronomy departments say they would like to end up in an academic 
position, while only 8\% express a desire to work in a national lab or 
research position (Mulvey \& Tesfaye 2006). By contrast, Figure~\ref{fig4} 
suggests that less than 50\% will ultimately obtain academic 
positions---and probably fewer, since the turnover at universities appears 
to be episodic. Graduate programs in astronomy should prepare their 
students for this reality.

Having documented the historical trends in the astronomy job market and 
the forces that help shape it, it should be straightforward to update the 
analysis periodically as new data emerge. In the future, I hope to make 
such updates available through the Career Services section of the AAS 
website\footnote{http://aas.org/career/}, and disseminate them through 
{\tt astro-ph}. With such information readily available, incoming graduate 
students can make informed decisions about their long-term career 
prospects, and faculty can nurture appropriate skills in the next 
generation of astronomers.

\acknowledgements 
This work was motivated by my involvement with the AAS Employment 
Committee, which is charged with facilitating the professional development 
and employment of astronomers at all career stages and on all career 
paths, and promoting balance and fairness in the job market. The National 
Center for Atmospheric Research is a federally funded research and 
development center sponsored by the National Science Foundation.


\end{document}